%Paper: hep-th/9204047
%From: Gordon Semenoff <semenoff@physics.ubc.ca>
%Date: 15 Apr 92 11:44 -0700

%%%%%%%%%%%%%%%%%%%%%%%%%%%%%%%%%%%%%%%%%%%%%%%%%%%%%%%%%%%%%%%%%%%%%%

%%%%%%%%%%%%%%%%%%%%% texfile,  no macros necessary %%%%%%%%%%%%%%%%%%

%%%%%%%%%%%%%%%%%%%%%%%%%%%%%%%%%%%%%%%%%%%%%%%%%%%%%%%%%%%%%%%%%%%%%%
\hfuzz 50pt
\magnification=1200

\centerline{\bf STRONG COUPLING QED BREAKS CHIRAL SYMMETRY}

\vskip 0.5truein
\centerline{\bf Gordon W. Semenoff}\footnote{}{This work is supported in
part by the Natural Sciences and Engineering Research Council of Canada.}
\vskip 0.4truein
\centerline{\it Department of Physics, University of British Columbia}

\centerline{\it Vancouver, British Columbia, Canada V6T 1Z1}
\vskip 0.5truein
\centerline{\bf Abstract}
\vskip 0.3truein
We show that the strong coupling limit of d-dimensional quantum
electrodynamics with $2^{d}/2^{[d/2]}$ flavors of fermions can be
mapped onto the s=1/2 quantum Heisenberg antiferromagnet in d-1 space
dimensions.  The staggered N\'eel order parameter is the expectation
value of a mass operator in QED and the spin-waves are pions.  We
speculate that the chiral symmetry breaking phase transition
corresponds to a transition between the flux phase and the
conventional N\'eel ordered phase of an antiferromagnetic t-J model.
\baselineskip=24pt
\vskip 1.0truein
The possibility that quantum electrodynamics (QED) can have a
nonperturbative ultraviolet fixed point has been investigated by many
authors [1-8].  Such a fixed point would give QED sensible ultraviolet
behavior and save it from triviality by avoiding the Landau ghost (or
Moscow zero) [9-10].  The most popular candidate for an ultraviolet
fixed point arises from a second order phase transition which occurs
when the electromagnetic coupling is increased to a critical value
where the theory breaks chiral symmetry.

The physical mechanism for such a phase transition advocated by
Miranksy [3] is that of `collapse' of the electron-positron
wave-function when the charge reaches a super-critical value, $e^2\sim
4\pi$.  There are two ways to screen super-critical charges.  In the
case of a super-critical nucleus the high electric field produces
electron-positron pairs, ejects the positron and absorbs the electron
to screen its charge.  Alternatively, the pair production is
suppressed by fermion mass, so the system can stabilize itself by
increasing the electron mass, thus the tendency to break chiral
symmetry.  These ideas are supported by studies of the Schwinger-Dyson
equations for QED in the quenched ladder approximation [3-5] which
find a line of fixed points in the $e^2$-plane between 0 and
$e_c^2=4\pi^2/3$ where the theory breaks chiral symmetry dynamically
and has a nontrivial continuum limit. This critical point has other
interesting behavior such as large negative anomalous dimensions for
fermion composite operators so that some four-fermion operators are
relevant [4]. There is also some numerical evidence
for this phase transition [7].

In this Letter we shall present a lattice model whose weak coupling
continuum limit is 4-flavor QED with light fermions and which, in its
strong coupling limit, exhibits spontaneously broken chiral symmetry.
Conventional analysis of Euclidean lattice
gauge theory formulates it as a classical statistical mechanics
problem and seeks second order phase transitions so that a nontrivial
continuum limit exists [11].  Here we formulate a Hamiltonian version
of lattice QED as quantum statistical mechanics [12].
In the latter a relativistic
continuum limit exists if there are gapless degrees of
freedom and if those degrees of freedom have a
relativistic dispersion relation, $\omega(k)\sim\vert k\vert$.  This
is, of course, true for the weak coupling continuum limit of the
lattice model which produces QED with 4 flavors of electron
(reminiscent of the standard model with four generations).  We show
that the strong coupling limit is equivalent to the
spin 1/2 quantum Heisenberg antiferromagnet and that the N\'eel order
coincides with chiral symmetry breaking.  In three or greater
dimensions, there is a rigorous proof that the ground state of the
s=1/2 antiferromagnet has N\'eel order [13].  Also, there is evidence from
numerical simulations and large spin expansions for N\'eel order in two
dimensions [14].  Furthermore, these systems exist in nature and can be
studied by experiment. In dimension 3 and higher the order persists
for some range of temperature.

This result indicates that in strong coupling 4-flavor QED has a
nontrivial continuum limit.  The
light excitations are the `pions' which coincide with the spin waves
of the antiferromagnet.  They have a relativistic dispersion relation
and interactions which are commonly represented by a
nonlinear sigma model.  Corrections to the strong coupling limit take
into account fermion hopping terms similar to those in a gauge
invariant t-J model. We conjecture that if $1/e^2$ is increased to
some critical value the chiral symmetry of QED is restored.  The
resulting phase of the t-J model is known as the flux phase which is a
plaquette-centered antiferromagnetic
gapless semiconductor rather than site-centered
antiferromagnetic insulator.

We shall use staggered fermions on a (d-1)-dimensional
lattice and continuum time which are obtained by spin-diagonalization
[15] of the naively latticized Dirac Hamiltonian $$
H_f={i\over2}\sum_{x,j}\left(\psi^{\dagger}(x)\alpha^j\nabla_j\psi(x)
-\nabla^j\psi^{\dagger}(x)\alpha^j\psi(x)\right)
=-{i\over2}\sum_{x,j}\left(\psi^{\dagger}(x+\hat j)\alpha^j\psi(x)-
\psi^{\dagger}(x)\alpha^j\psi(x+\hat j)\right)
\eqno(1)
$$ where $\nabla^j$ is the forward lattice difference operator,
$\nabla_jf(x)=f(x+\hat j)-f(x)$, $\alpha^j$ are the $2^{[d/2]}\times
2^{[d/2]}$ Hermitean Dirac matrices, $x,y,\dots$ refer to sites on a
hypercubic lattice, and $\hat i, \hat j,\dots$ refer to unit vectors.
(Here $[d/2]$ is the largest integer $\leq d/2$.)  The second form of
the Hamiltonian in (1) describes a fermion hopping problem in a
U($2^{[d/2]}$) background gauge field given by the unitary matrices
$\alpha^j$.  The crucial observation which allows spin diagonalization
is that this background field has only U(1) curvature, i.e. if we
consider the product around any plaquette
$\alpha^j\alpha^k(\alpha^j)^{\dagger}(\alpha^k)^{\dagger}=-1$.  This
allows diagonalization using the gauge transformation
$\psi(x)\rightarrow
(\alpha^1)^{x_1}(\alpha^2)^{x_2}\dots(\alpha^{(d-1)})^{x_{(d-1)}}\psi(x)$
resulting in the Hamiltonian $$
H_f=-{i\over2}\sum_{x,j}(-1)^{\sum_1^{j-1}x_p}\left(\psi^{\dagger}(x+\hat
j)\psi(x)-\psi^{\dagger}(x)\psi(x+\hat j)\right)
\eqno(2)
$$ which describes $2^{[d/2]}$ species of lattice fermions with
background U(1) magnetic flux $\pi$ through every plaquette of the
lattice. Each species of fermion must have the same spectrum as the
original one given by the Dirac Hamiltonian (1). This allows reduction
of the fermion multiplicity by a factor of $2^{[d/2]}$.  The result
resembles a condensed matter hopping problem with a single species of
fermion where there is a background magnetic field $\pi$ per
plaquette.

Chiral symmetries are obtained by lattice translations by
one site. This translation interchanges the even ($\sum_1^{d-1}
x_p$=even) and odd ($\sum_1^{d-1}x_p=$odd) sublattices.  The
substitutions $$
\psi(x)\rightarrow (-1)^{\sum_{j+1}^{d-1}x_p}\psi(x+\hat j)
\eqno(3)
$$ leaves the Hamiltonian in (2) invariant.  A candidate for Dirac
mass operator, which changes sign under the transformations in (3), is
the staggered charge density operator $$
\mu=\sum_x (-1)^{\sum_1^{d-1}x_p}\psi^{\dagger}(x)\psi(x)
\eqno(4)
$$ If we introduce N species of lattice fermions, the continuum limit
of (2) describes $2^{d-1}N/2^{[d/2]}$ species of massless Dirac fermions.

To obtain the continuum limit and the number of fermion species, we
first divide the lattice into $2^{d-1}$ sublattices according to
whether the components of their coordinates are even or odd.  For
example, when (d-1)=3, we label 8 fermion species as $\psi({\rm
even},{\rm even},{\rm even})\equiv\psi_1$, $\psi({\rm even},{\rm
odd},{\rm odd})\equiv\psi_2$, $\psi({\rm odd},{\rm even},{\rm
odd})\equiv\psi_3$, $\psi({\rm odd},{\rm odd},{\rm
even})\equiv\psi_4$, $\psi({\rm even},{\rm even},{\rm
odd})\equiv\psi_5$, $\psi({\rm even},{\rm odd},{\rm
even})\equiv\psi_6$, $\psi({\rm odd},{\rm even},{\rm
even})\equiv\psi_7$, $\psi({\rm odd},{\rm odd},{\rm
odd})\equiv\psi_8$.  Then, if we add the mass operator in (4), in
momentum space the Hamiltonian has the form $$
H_f=\int_{\Omega_B}d^3k~\psi^{\dagger}(k)\left(A^i\sin k_i +B
m\right)\psi(k)
\eqno(5)
$$ where the $8\times8$ Dirac matrices are
$\psi(k)=(\psi_1,\dots,\psi_8)$, $A^i=\left(\matrix{0&\alpha^i\cr
\alpha^i&0\cr}\right)$, $B=\left(\matrix{1&0\cr0&-1\cr}\right)$, $$
\alpha^1=\left(\matrix{0&1\cr 1&0\cr}\right)
{}~~,~~
\alpha^2=\left(\matrix{\sigma^1&0\cr 0&-\sigma^1\cr}\right)
{}~~,~~
\alpha^3=\left(\matrix{\sigma^3&0\cr 0&-\sigma^3\cr}\right)
\eqno(6)
$$ $\sigma^i$ are Pauli matrices, we have used the Fourier transform
$\psi(x)=\int_{\Omega_B}{d^3k\over(2\pi)^{3/2}}e^{-ik\cdot x}\psi(k)$
and $\Omega_B$ is the Brillouin zone of the (even,even,even)
sublattice, $-\pi/2<k_i\leq\pi/2$. The fermion spectrum is
$\omega(k)=\sqrt{\sum_i\sin^2k_i+m^2}$ and only the region $k_i\sim0$
is relevant to the continuum limit. We have normalized $\psi(k)$ so
that
$$
\left\{\psi(x),\psi^{\dagger}(y)\right\}=\delta(x-y)
{}~~,~~
\left\{\psi(k),\psi^{\dagger}(l)\right\}=\delta(k-l)
\eqno(7)
$$ If we define $\beta=\left(\matrix{\sigma^2&0\cr0&-\sigma^2}\right)$
and the unitary matrix
$M={1\over2}\left(\matrix{1-\beta&1+\beta\cr1+\beta&1-\beta\cr}\right)$
and $\psi=M\psi'$ with $\psi'=(\psi_a,\psi_b)$ the Hamiltonian is $$
H_f=\int_{\Omega_B}d^3k~\left(\psi_a^{\dagger},\psi_b^{\dagger}\right)
\left(\matrix{\alpha^i\sin k_i-\beta m&0\cr0&\alpha^i\sin k_i+\beta
m\cr}\right)\left(\matrix{\psi_a\cr\psi_b}\right)
\eqno(8)
$$ In the low momentum limit, $\sin k_i\sim k_i$, with fermion density
1/2 per site, we obtain 2 continuum Dirac fermions.  Furthermore, the
staggered charge operator gives a Dirac mass of differing sign for the
two species.  In d dimensions the proceedure is similar to this. In
general, we shall consider $N$ lattice species in d dimensions which
yeilds $2^{(d-1)}N/2^{[d/2]}$ continuum species of Dirac fermions
where the lattice fermion density is N/2 per site.

In lattice electrodynamics the gauge field $A_i(x)$ and electric
field $E_i(x)$ associated with the link between $x$ and $x+\hat i$ are
conjugate variables, $
\bigl[ A_i(x), E_i(y)\bigr]=ie^2\delta_{ij}\delta(x-y)
$. We shall represent this commutator by taking the quantum states as
functions of $A_i(x)$ and $E_i(x)=-ie^2\partial/\partial A_i(x)$.  The
anticommutator algebra (7) is represented by a $2^N$-level fermion
system at each site with cyclic vector
defined by $\psi(x)\vert 0>=0~\forall x$.  The Hamiltonian is $$
H=\sum_{x,j}-{e^2\over2}{\partial^2\over\partial
A_j(x)^2}+\sum_{x,ij}{1\over2e^2}F_{ij}^2
+\sum_{x,j}\left(t_{x,j}\psi^{\dagger}(x+\hat
j)e^{iA_j(x)}\psi(x)+h.c.\right)
\eqno(9)$$
where $F_{ij}=\nabla_iA_j-\nabla_jA_i$ for noncompact QED,
$F_{ij}=\sin(\nabla_iA_j-\nabla_jA_i)$ for compact QED, and $t_{x,i}$
contains a background field of $\pi$ (mod $2\pi$) through every
plaquette of the lattice.  The Hamiltonian is invariant
under the gauge transformation $A_i(x)\rightarrow
A_i(x)+\nabla_i\chi(x)$, $\psi(x)\rightarrow
e^{i\chi(x)}\psi(x)$ which is generated by
$$ {\cal G}(x)=\sum_j\nabla_j {1\over
i}{\partial\over\partial A_j(x-\hat j)}+
\psi^{\dagger}(x)\psi(x)-N/2
\eqno(10)
$$ Gauge invariance is imposed as a physical state condition, ${\cal
G}(x)\Psi_{\rm phys}(A)=0$.

We shall first consider the case $N=2$.  Then there are 4 flavors of 4
component fermions in 4 dimensions and 2 flavors of 4 component
fermions in 2+1 dimensions.  For strong coupling perturbation theory
we write the Hamiltonian as, $H=H_0+H_1+H_2$ where
$H_0=-{e^2\over2}\sum{\partial^2 \over\partial A^2}$,
$H_1=\sum t\psi^{\dagger}
\psi+h.c.$ and $H_2={1\over4e^2}\sum F_{ij}^2$ are each
gauge invariant operators.  The leading order
ground state is the ground state of
$H_0$ which is $A$-independent and is gauge
invariant when each state has
fermion occupation number one,
$
\vert\Psi_0[\tau_x])=\prod_x\psi_{\tau_x}(x) \vert0>
$.
This is a normalizable state in
compact QED where $A_i(x)$ is integrated from 0 to $2\pi$ and is a
non-normalizable component of a continuum spectrum in non-compact
QED.  It is highly degenerate: each of the $2^V$
(where $V$ is volume) components labelled by $[\tau_x]$ has
the same eigenvalue of $H_0$.
The degeneracy is resolved by
diagonalizing the matrix elements of perturbations in $1/e^2$.  First order
perturbations to the vacuum energy vanish.  Thus, the leading term in
the vacuum energy is of order $1/e^2$, and is given by the lowest
eigenvalue of the matrix $$
\delta_2E_0=-(\Psi_0[\tau_x]\vert H_1{1\over H_0-E_0}H_1\vert\Psi_0[\tau_x'])
+(\Psi_0[\tau_x]\vert H_2\vert\Psi_0[\tau_x'])
\eqno(11)
$$ The second term is diagonal and therefore is irrelevent to
resolving the degeneracy.  Diagonalizing the matrix in the first term
is equivalent to solving the eigenvalue problem for the four-fermion
Hamiltonian
$$
H_{\rm eff}=-{4\over e^2}\sum_{x,i}\vert t_{x,i}\vert^2
\psi^{\dagger}(x+ \hat i)\psi(x)\psi^{\dagger}(x)\psi(x+\hat i)
\eqno(12)
$$ restricted to the subspace of the Hilbert space where each site is
singly occupied.  With the identity
$-2\psi^{\dagger}(x)\psi(y)\psi^{\dagger}(y)\psi(x)=
\psi^{\dagger}(x)\vec\sigma\psi(x)\cdot\psi^{\dagger}(y)
\vec\sigma\psi(y)+\psi^{\dagger}(x)\psi(x)\psi^{\dagger}(y)\psi(y)$
we obtain the Hamiltonian of the s=1/2 quantum antiferromagnet, $$
H_{\rm eff}={2\over e^2}\sum_{x,i}\vert t_{x,i}\vert^2
\psi^{\dagger}(x+\hat i)\vec\sigma\psi(x+ \hat
i)\cdot\psi^{\dagger}(x)\vec\sigma\psi(x)+{\rm ~const.}
\eqno(13)
$$

Fermion mass operators are staggered charge density operators.  For
N=2, consider a mass operator with differing signs for the two lattice
flavors, i.e.  $m\sum_x (-1)^{\sum x}\psi^{\dagger}\sigma^3\psi$ which
in the naive continuum limit corresponds to the mass matrix
$\left(\matrix{m&0&0&0\cr0&-m&0&0\cr0&0&-m&0\cr0&0&0&m\cr}\right)$.
(Note that there is no chiral anomaly in this channel.) This
corresponds to a staggered magnetization operator and its expectation
value is the order parameter of the antiferromagnet.  There is a proof
that for s=1/2 and (d-1)$\geq3$, (and it is widely believed to also
hold in d-1=2 although there is no rigorous proof) that the Heisenberg
antiferromagnet has a N\'eel ordered ground state, i.e.  $$
\lim_{m\rightarrow0}\lim_{V\rightarrow\infty}<\sum_x
(-1)^{\sum x}\psi^{\dagger}\sigma^3\psi>\neq 0
\eqno(14)
$$ In quantum electrodynamics, this is the chiral limit and implies
that there is a nontrivial expectation value of the mass operator and
thus spontaneous chiral symmetry breaking. In (d-1)$\geq3$ this
persists for some finite range of temperature whereas in
(d-1)=2 it can only be
true at zero temperature.

We conclude that, in the infinite coupling limit, $2^d/2^{[d/2]}$
flavor (4 flavor in d=4) QED breaks chiral symmetry. The only light
particles in the spectrum are spin-waves which are the Goldsone bosons
for breaking of the SU(2) symmetry of the Heisenberg model.  In QED
they are the `pions' corresponding to the breaking of the chiral
symmetry and electrons and all other charged excitations are confined.
There are not enough spin waves to account for the
breaking of $SU_L(4)\times SU_R(4)$ chiral symmetry.
The reason for this mismatch of the
number of Goldstone bosons and broken symmetries arises from the
presence of relevant operators which reduce some of the
apparent continuous chiral symmetries of the naive continuum theory to
discrete ones of the lattice.  These discrete symmetries are broken
but do not require Goldstone bosons.  The only true flavor symmetry of the
lattice theory is SU(2).  The apparent $SU_L(4)\times SU_R(4)$
symmetry of the weak
coupling limit stems from the fact that all of the operators
which could break $SU_L(4)\times SU_R(4)$ to SU(2) plus
discrete chiral symmetries are
irrelevant in that limit.  However, they can become relevant at strong
coupling and the full $SU_L(4)\times SU_R(4)$ symmetry is absent.
Also, recall that
Goldstone's theorem applies to the continuous symmetries of the bare
Hamiltonian.  The effective Hamiltonian (13) is invariant under the
discrete chiral transformations in (3).  However, the ordered ground
state is not - therefore the discrete chiral symmetries which
involve translations by one lattice site are also
broken as $1/e^2\rightarrow0$.

We can think of the higher order in $1/e^2$ corrections as
perturbations of the
Heisenberg model.  When we increase $1/e^2$ we eventually should
arrive at a critical coupling where the chiral symmetry is restored.
A future technical problem will be to estimate the critical coupling.
Also, spin-wave analysis which is exceptionally good for the quantum
antiferromagnet can also be used to look at the dynamics of
strong-coupling QED.

We expect that chiral symmetry breaking should persist for $N$ in the
vicinity of $N=2$.  However, when N=1 (and whenever N is odd), since
operator $\psi^{\dagger}\psi$ has integer eigenvalues, it is not
possible to find states which are annihilated by the gauge generator
in (10) without nontrivial electric fields.  The strong coupling
ground state energy is necessarily of order $e^2$ and the ground state
is the lowest eigenstate of the effective coulomb Hamiltonian, $$
H_{c}=\sum_x{e^2\over2}(\psi^{\dagger}\psi-1/2){1\over-\nabla^2}
(\psi^{\dagger}\psi-1/2)
\eqno(15)
$$ We expect that the ground state of $H_c$ is a chiral symmetry
breaking Wigner lattice where either the even or odd sublattice is
completely occupied and the other sublattice is completely empty.
Thus, we expect chiral symmetry breaking in this case too, with
critical behavior in the universality class of the Ising model.  However,
we point out that the vacuum energy is not a smooth function of $N$,
being of order $e^2$ when N is odd and of order $1/e^2$ when N is
even.

In 2+1-dimensions, our results suggest chiral symmetry breaking,
at least for N=1 and 2, in agreement with the analysis of
quenched Schwinger-Dyson equations [16].  It is also known
that at large $N$ other phases, such as the flux phase itself
[17] compete with the N\'eel ordered phase of the antiferromagnet.
It would be interesting to examine this further, in particular to
obtain an estimate of upper critical $N$, if one exists, in the
context of the present model.

Salmhofer and Seiler [8] have given a proof that QED defined on a
spacetime lattice with $\geq 4$ flavors and in $\geq 4$ dimensions
breaks chiral symmetry.  Their model differs from ours in the order
of symmetry of the strong coupling versus the weak coupling limit.
Their 4-flavor QED has only $Z_2$ symmetry at strong coupling and
no Goldstone bosons.  Furthermore, the order parameters differ in
the two cases, theirs being in the chiral U(1) channel.  It is also
a puzzle to us that the vacuum energy of their model in the large
$e^2$ limit is of order one, whereas we find that it is always either
of order $e^2$ or $1/e^2$ depending on whether N is even or odd.
Resolution of these differences is an important problem.

\centerline{\bf References}
\item{1.}M. Gell Mann and F. Low, Phys. Rev. 95 (1954), 1300.
\item{2.}K. Johnson, M. Baker and R. Willey, Phys. Rev. Lett. 11
(1963), 518; Phys. Rev. 136 (1964), 1964.
\item{3.}V. Miransky, Nuov. Cim. 90A (1985), 149.
\item{4.}C. Leung, S. Love and W. Bardeen, Phys. Rev. Lett. 56
(1986), 1230; Nucl. Phys. B273 (1986), 649.
\item{5.}K. Yamawaki, M. Bando and K. Matumoto, Phys. Rev. Lett. 56
(1986), 1335.
\item{6.}J. Kogut and E. Dagotto, Phys. Rev. Lett. 59 (1987), 617.
\item{7.}J. Kogut, E. Dagotto and A. Kocic, Phys. Rev. Lett. 60
(1988), 772.
\item{8.}M. Salmhofer and E. Seiler, Comm. Math. Phys. 139 (1991), 395.
\item{9.}L. Landau and I. Pomeranchuk, Dokl. Acad. Nauk. SSSR, 102 (1955),
489; L. Landau, in {\it Neils Bohr and the Development of Physics},
ed. W. Pauli (Pergamon, London, 1955).
\item{10.}E. Fradkin, JETP 28 (1955), 750.
\item{11.}K. Wilson, Phys. Rev. D20 (1974), 2445.
\item{12.}For a review of Hamiltonian methods, see
J. Kogut, Rev. Mod. Phys. 55 (1983), 775.
\item{13.}T. Kennedy, E. Lieb and B. Shastry, J. Stat. Phys. 53
(1988), 1019.
\item{14.}For a recent review, see E. Manousakis, Rev. Mod. Phys. 63
(1991), 1.
\item{15.}H. Kluberg-Stern, A. Morel, O. Napoly and B. Petersson, Nucl.
Phys. B220[FS8] (1983), 447.
\item{16.}T. Applequist, D. Nash and L. C. R. Wijewardhana, Phys.
Rev. Lett. 60 (1988), 2575; D. Nash, Phys. Rev. Lett. 62 (1989), 3024.
\item{17.}I. Affleck and B. Marston, Phys. Rev. B37 (19888), 3774.

 \end